\documentstyle[prl,aps,epsf]{revtex} \def\narrowtext{} \tighten \twocolumn
\input epsf.sty

\begin{document}
\draft

\title{
The neutron resonance: modeling photoemission and tunneling data in the
superconducting state of Bi$_2$Sr$_2$CaCu$_2$O$_{8+\delta}$}
\author{M. Eschrig and M. R. Norman}
\address{ Materials Science Division, Argonne National Laboratory, Argonne,
Illinois 60439}
\address{%
\begin{minipage}[t]{6.0in}
\begin{abstract}
Motivated by neutron scattering data, we develop
a model of electrons interacting with a magnetic resonance and use it to
analyze angle resolved photoemission (ARPES) and tunneling data
in the superconducting state of Bi$_2$Sr$_2$CaCu$_2$O$_{8+\delta}$.
We not only can explain the peak-dip-hump structure observed
near the $(\pi,0)$ point, and its particle-hole asymmetry as seen in SIN
tunneling spectra,
but also its evolution throughout the Brillouin zone, including a velocity
`kink' near the d-wave node.
\typeout{polish abstract}
\end{abstract}
\end{minipage}}

\maketitle
\narrowtext

Recent advances in the momentum resolution of ARPES spectroscopy have led to a
detailed mapping of the spectral lineshape in the high-T$_c$ superconductor
Bi${_2}$Sr${_2}$CaCu${_2}$O${_{8+\delta}}$ (BSCCO) throughout the Brillouin
zone.\cite{Kaminski00,Bogdanov00}
In contrast to normal state data where well defined excitations do
not exist, quasiparticle peaks were identified
below $T_c$ along the entire Fermi surface.\cite{Kaminski99}
Moreover, it has been known for some time that near the
$(\pi,0)$ ($M$) point of the zone, the spectral function
shows an anomalous lineshape, the so called `peak-dip-hump' 
structure.\cite{Dessau91,Ding96}
The new data indicate
a) near the d-wave node of the superconducting gap, the dispersion
shows a characteristic `kink' feature: for $|\omega| < \omega_{kink}$,
the spectra exhibit sharp peaks with a weaker dispersion;
above this, broad peaks with a stronger dispersion;\cite{Kaminski00,Bogdanov00} 
b) away from the node, the dispersion
kink develops into a `break'; the two resulting
branches are separated by an energy gap, and overlap in momentum space;
c) towards $M$, the break evolves into a pronounced spectral `dip' separating
the almost dispersionless quasiparticle branch from the weakly dispersing
high energy branch (the `hump');
d) the kink, break, and dip features all occur at the same energy,
independent of position in the zone.\cite{Kaminski00}

Features similar to the ARPES spectrum near the $M$ point
were earlier observed in tunneling spectroscopy.\cite{Huang89}
Experimental SIN junctions on BSCCO show a 
characteristic asymmetry, with a 
more pronounced dip-hump structure on the occupied side.\cite{Renner95}
On the other hand, SIS junctions reveal a strong dip-hump feature on both
bias sides.\cite{Mandrus91}

There have been several theoretical treatments which assigned the anomalous
ARPES lineshape near the $M$ point of the zone
to the coupling between spin fluctuations and 
electrons.\cite{Dahm96,Shen97,Norman97,Abanov99} 
Here, we are able to explain features a)-d) of the ARPES data,
as well as the SIN tunneling asymmetry,
in terms of the combined effect 
of A) the flat electronic dispersion near the $M$ point of the zone
and B) coupling of the fermionic degrees of freedom to a bosonic 
mode which is sharp in energy and peaked in momentum near
$\vec{Q}=(\pi,\pi)$.
Our main result is that the anomalous features in the dispersion and lineshape
for all points in the zone have the same origin.

A resonance mode with these characteristics
is observed in inelastic neutron scattering experiments 
in bilayer cuprates in the
superconducting state.\cite{Rossat91,Fong99}
The neutron resonance lies below a gapped continuum,
the latter having a signal typically
a factor of 30 less than the maximum at $\vec{Q}$ at the mode 
energy.\cite{Bourges97}
In order to extract the essential physics, we concentrate on the
mode part and neglect the continuum.
The latter contributes mainly to additional damping at higher energies.
We treat the mode in a semi-phenomenological way, taking the relevant 
parameters from experiment.
We then calculate the resulting electronic self energy
to second order in the coupling constant.
 From the self energy we directly obtain the
spectral function measured by ARPES, which we then use
to calculate the tunneling conductance.

The retarded self energy on the real energy axis is given by\cite{Rammer86}
\begin{equation}
\hat\Sigma^R= -\frac{ig^2}{2\mu_B^2} \hat\tau_3
\left( \hat G^K * \chi^R+\hat G^R *\chi^K\right) \hat\tau_3
\end{equation}
with 
$A*B(\vec{k},\epsilon)\equiv \sum_{\vec{q}} \int_{-\infty}^\infty
\frac{d\omega}{2\pi}
A(\vec{k}-\vec{q},\epsilon-\omega) B(\vec{q},\omega)$, $\hat \tau_i$ Pauli
matrices in particle-hole space, and $g$ the effective coupling constant. 
The Keldysh ($K$) components are given in terms of retarded ($R$)
and advanced ($A$) functions by
$\hat G^K=(\hat G^R-\hat G^A)(1-2f)$ and $\chi^K=(\chi^R-\chi^A)(1+2b)$,
with the usual Bose ($b$) and Fermi ($f$) distribution functions.

The model for the mode is based on measurements of the 
spin susceptibility from inelastic neutron scattering experiments.\cite{Fong99}
The mode energy will be denoted by $\Omega $ and its energy 
width is assumed to be 
irrelevant for the self energy. This assumption will be confirmed
later.
This leads to the following model
for the mode part of the susceptibility 
\begin{equation}
\chi^{R/A} (\vec{q},\omega) = -f(\vec{q}) \left( \frac{1}{\omega-\Omega
\pm i \delta } - \frac{1}{\omega+\Omega \pm i \delta} \right)
\end{equation}
Here $f(\vec{q})$ describes the momentum dependence of the mode and is assumed
to
be enhanced at the $(\pi,\pi)$ point. Using the correlation length $\xi$ we
write it as
\begin{equation}
f(\vec{q})=\frac{\chi_{\vec{Q}}
\xi^{-2}}{\xi^{-2}+4 (\cos^2\frac{q_x}{2}+\cos^2\frac{q_y}{2})}
\end{equation}
Experimentally the energy integrated susceptibility at the $(\pi, \pi )$
wavevector, $\pi \chi_{\vec{Q}}$, was determined to be $0.95 \mu_B^2$ per plane
for BSCCO,\cite{Fong99} leading
to $\chi_{\vec{Q}} =0.3 \mu_B^2$.
For the correlation length, we take a conservative estimate of $\xi=2a$.
This corresponds to a
full width half maximum of $0.26 $\AA$^{-1}$, as observed in YBCO, but
somewhat smaller than that estimated for BSCCO ($0.52 $\AA$^{-1}$)\cite{Fong99}
which we feel is somewhat broad.
The mode energy was chosen to be $\Omega=39$ meV, which represents the
reported values between 35 and 43 meV.\cite{Fong99}

Though $\chi$ is `renormalized', we use a bare $\hat G$ in Eq.~1.
This is the same approximation as in Ref.~\onlinecite{Vilk97}, where it
was shown that this is better than using renormalized
$\hat G$ with vertex corrections neglected.  This is unlike the
electron-phonon problem, where Migdal's theorem applies.
We take the success of explaining the experimental
features as strong support of this approximation. 

The bare Green's functions with normal state dispersion
$\xi_{\vec{k}}$, gap function $\Delta_{\vec{k}}$, and excitation energy
$E_{\vec{k}}=\sqrt{\xi_{\vec{k}}^2+\Delta_{\vec{k}}^2}$ are
\begin{equation}
\hat G^{R/A} (\vec{k},\epsilon ) = \frac{\hat \alpha_{\vec{k}}}
{\epsilon -E_{\vec{k}} \pm i \Gamma } +
\frac{\hat \beta_{\vec{k}}}{\epsilon+E_{\vec{k}}\pm i \Gamma }
\end{equation}
where $\alpha_{11}=(1+\xi_{\vec{k}}/E_{\vec{k}})/2$, 
$\beta_{11}=(1-\xi_{\vec{k}}/E_{\vec{k}})/2$,
$\alpha_{12}=-\beta_{12}=-\Delta_{\vec{k}}/2E_{\vec{k}}$, etc.
For the normal state dispersion we use a 
six-parameter tight binding fit.\cite{Norman95} 
We neglect bilayer splitting, as experiments suggest it is
absent in BSCCO.\cite{Ding96} 
A characteristic feature of this dispersion is a flat band
with a saddlepoint at $M$ with
energy $\xi_M=-$34 meV.
The superconducting gap function is taken to be the d-wave
$\Delta_{\vec{k}}=\Delta_0 (\cos k_x-\cos k_y)/2$ with a maximal gap value of
$\Delta_0=35 $ meV. We have chosen $\Gamma=5$ meV as an intrinsic
lifetime broadening. 
The coupling constant relevant for our model is $g^2\chi_{\vec{Q}}$,
chosen to be 0.125 $\mbox{eV}^2 \mu_B^2$. 
Given a value $\chi_{\vec{Q}}=0.3 \mu_B^2$,
this corresponds to $g=0.65 $ eV, the same value as used in previous
spin fluctuation work.\cite{Monthoux94}
We performed the $\omega $-integration in Eq.~1 analytically and
the correlation product in momentum space via fast Fourier transform,
using $256\times 256$ points in the Brillouin zone.

In Fig.~1 we show the renormalization function
$Z(\epsilon) = 1-\Sigma_0(\epsilon )/\epsilon$, where $\Sigma_0$ is
the $\hat \tau_0$ component of the $\hat \Sigma$ matrix.
Since the $\vec{q}$ integral in Eq.~1 is dominated
by the regions around the $M$
point where the band is flat and close to the chemical potential, there
are features in the imaginary part of the self-energy
connected with the two extremal energies
$\Delta_0$ and $E_M=\sqrt{\xi_M^2+\Delta_0^2}$.
These features do not show dispersion, but a change in magnitude with
position in the zone which is determined by the momentum width of the mode. 
This is the central result of this paper.
More generally, the imaginary part of the self energy is enhanced between
the values 
\begin{figure}
\epsfxsize=3.4in
\epsfbox{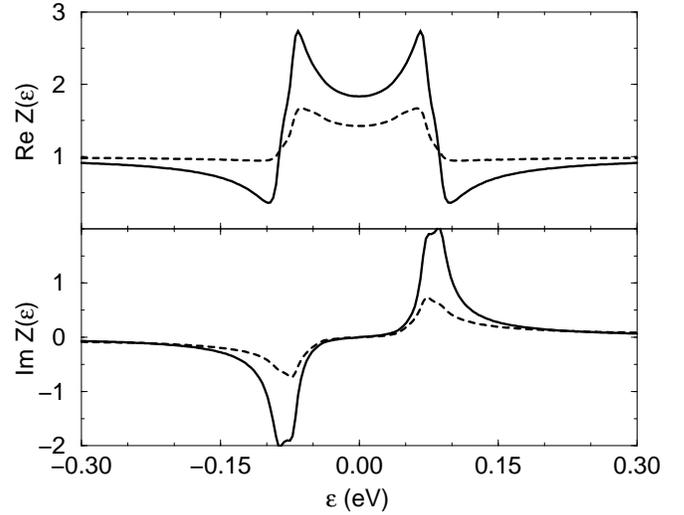}
\caption{ \label{fig1}
Renormalization function $Z(\epsilon )$ at the $M$=$(\pi,0)$ point 
(solid line) and at the node $(0.36\pi,0.36\pi)$ (dashed line) for $T=40$K.
}
\end{figure}
\noindent
$\epsilon_1=\Delta_0+\Omega$ and $\epsilon_2= E_M+\Omega $. 
For $\xi_M$ approaching the chemical potential,
$E_M $ approaches $\Delta_0$ resulting
in a peak-like feature in the self energy.
For our case, $\epsilon_1=74$ meV and $\epsilon_2=88 $ meV.
Because the spectral weight of 
the mode is maximal near 
$\vec{Q}=(\pi,\pi)$,
the $M$ points of the
zone, which are connected by $\vec{Q}$, benefit mostly. This results
in stronger features in the self energy near the $M$ points
compared to e.g.~the nodal points.
The peaked structure in the imaginary part of the self energy results 
(via Kramers-Kronig relations)
in an 
enhancement of the real part of the renormalization function for $|\epsilon |
< \epsilon_1$, and a reduction of it for $|\epsilon | > \epsilon_2$, 
as shown in Fig.~1. 
This leads to a renormalization of the
low-energy dispersion of the spectra compared to the high energy part.
Since the experimental energy width of the neutron resonance
is smaller than the variation in energy 
of typical features in the self energy, 
this confirms our assumption that the energy width of the mode is
not relevant.

The spectral function is obtained by
\begin{equation}
A(\vec{k},\epsilon) = -2 \mbox{Im} \left[ \left( 
\hat G^{R}(\vec{k},\epsilon)^{-1} -\hat \Sigma^{R} (\vec{k},\epsilon )
\right)^{-1} \right]_{11}
\end{equation}
In Fig.~2 we show the spectral functions for momentum cuts through
the $M$ point towards $Y=(\pi, \pi )$ ($MY$ cut),
and parallel to $MY$ through the nodal point. Due to particle-hole coherence
factors, there are quasiparticle peaks at $M$ on both sides of
the chemical potential.
On the negative energy side, the peak is more pronounced since $\xi_M$ is
negative, and a strong dip feature is present.
The asymmetry of the dip feature is a combined effect of the $\hat \tau_3$
component of $\hat \Sigma$, which introduces particle-hole asymmetry, 
and the inherent particle-hole asymmetry of the band structure near the
$M$ point. Going from $M$ towards the Fermi surface (Fig.~2, bottom),
the hump feature quickly loses weight as
observed in ARPES.\cite{Kaminski00}
\begin{figure}
\epsfxsize=3.4in
\epsfbox{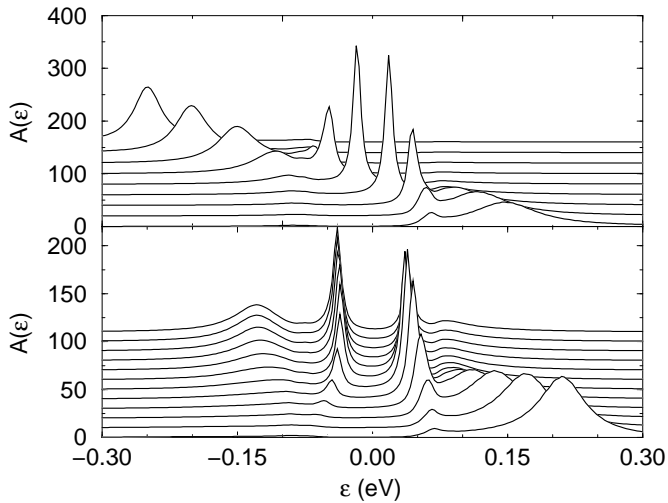}
\caption{ \label{fig2}
Spectral function $A(\vec{k}, \epsilon )$ for $T=40$K. 
Top panel:
for $\vec{k}$ points
along the cut parallel to $MY$ through the order parameter node, 
from $(0.20 \pi, 0.36 \pi )$ to $(0.51 \pi, 0.36 \pi )$.
Bottom panel: for $\vec{k}$ points along the $MY$ cut from
$(0, \pi)$ to $(0.43 \pi, \pi )$.
}
\end{figure}
\noindent
In the top panel of Fig.~2 we show spectra for a cut parallel to
$MY$ through the order parameter node at the Fermi surface.
Near the node there is only one peak crossing
the chemical potential. The dip-hump features are 
very weak near the node and are
presumably overshadowed by the additional lifetime effects due to the continuum 
part of the spin susceptibility. Note the much broader peaks for higher
energies, $|\epsilon| > 80 $ meV, compared to the sharper peaks near
the chemical potential, as observed in ARPES
experiments.\cite{Kaminski00,Bogdanov00,Kaminski99}

In Fig.~3 we present our results for the dispersion
obtained from the maxima of the occupied part of the spectral function,
$A(\vec{k},\epsilon) f(\epsilon)$. 
Near the $M$ point we observe an
almost dispersionless strong peak feature at roughly the gap energy
$-\Delta_0$, and a weaker hump feature at slightly below
$-\epsilon_2$, consistent with experimental finding c). 
The peak feature, which without interaction with the mode would be at
$E_M$, is pushed towards the chemical potential,
thus ending up close to $\Delta_0$ for not too small coupling 
constants. The position of the hump feature is strongly dependent on
the coupling constant.  
We adjusted $g$ to reproduce the experimental value
of about -130 meV for the hump feature at $M$;
this choice also results in the weak dispersion of the hump feature as
observed in experiment.\cite{Kaminski00}
As one goes away from $M$, the dispersion of the hump
extents further below $-\epsilon_2$ and the peak starts to show dispersion, 
until a characteristic break in the dispersion with a 
jump at $\sim$-80 meV develops. This is exactly the experimental finding
in point b). Note the stability of the characteristic
-80 meV energy value for the break/dip feature throughout the zone.
This is a result of the dominance of the region near $M$ in the $\vec{q}$
sum in Eq.~1, which
sets the energy scale. Thus we confirm point d) of the experimental findings.
\begin{figure}
\epsfxsize=3.1in
\epsfbox{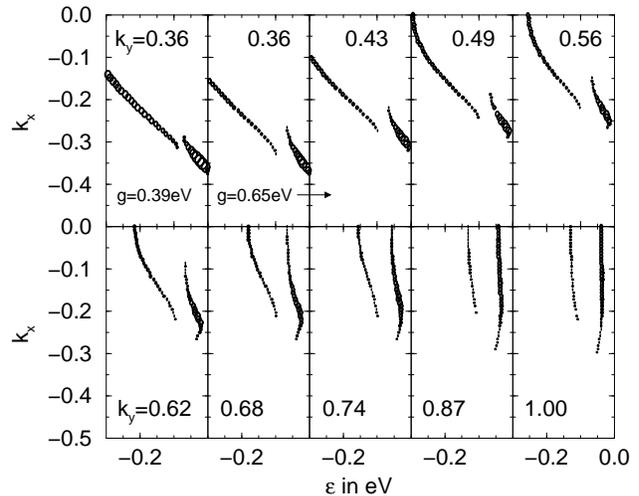}
\caption{ \label{fig3}
Calculated dispersion at $T=40$K obtained from the maxima of the occupied part
of the spectral function for cuts along $k_x$ for fixed $k_y$ as indicated.
Momenta are in units of $\pi $. The size of the symbols represents the
peak intensity. The coupling constant is $g=0.65 $ eV except in panel 1, where
we show for comparison with panel 2 results for $g=0.39 $ eV.
Note striking similarity of this plot to the data of
Ref.~\protect\onlinecite{Kaminski00}.
}
\end{figure}
\noindent
As the nodal point is approached, the self energy
becomes weaker due to the momentum dependence of the mode.
The sudden change in the linewidth 
for a cut parallel to $MY$ through the node 
(panel 2 of Fig.~3), as discussed in Fig.~2, 
occurs around -80 meV, in accordance with point a).
We still observe a weak break feature, which will be smeared out by
additional lifetime broadening from the continuum part of the
susceptibility.  This weak break
is also reduced for a smaller coupling (panel 1), or if the Lorentzian
in Eq.~3 is replaced by a Gaussian.
Note that in accordance with 
experiments, the velocity near the nodal point is reduced 
compared to that for higher binding energies, causing a velocity `kink'.

Knowing the spectral function throughout the zone,
we are able to calculate the tunneling spectra
given a tunneling matrix element $T_{\vec{k}\vec{p}}$.
 From the SIN tunneling current $I(V)$
one obtains the differential conductance, $dI/dV$.
As usual we neglect the energy dependence of the SIN matrix element
$|M_{\vec{k} }|^2 = 2e\sum_{\vec{p}}|T_{\vec{k}\vec{p}}|^2
A_N(\vec{p}, \epsilon )$, where $A_N$ is the spectral function of the normal
metal. The SIN tunneling current is then given by
\begin{eqnarray}
I(V)=\sum_{\vec{k} } |M_{\vec{k}}|^2
\int_{-\infty}^\infty \frac{d\epsilon }{2\pi }
A(\vec{k},\epsilon ) 
\left\{ f(\epsilon ) - f(\epsilon + eV) \right\}
\end{eqnarray}
In the top panels of Fig.~4, we show results for the
SIN $dI/dV$ for several coupling strengths.
We model the tunneling matrix element for two extreme cases: for
incoherent tunneling we assume a constant
$|M_{\vec{k} }|^2 =M_0^2$, 
whereas for coherent tunneling we use
$|M_{\vec{k} }|^2=\frac{1}{4}M_1^2( \cos k_x - \cos k_y )^2$.\cite{Chak93}
\begin{figure}
\epsfxsize=3.4in
\epsfbox{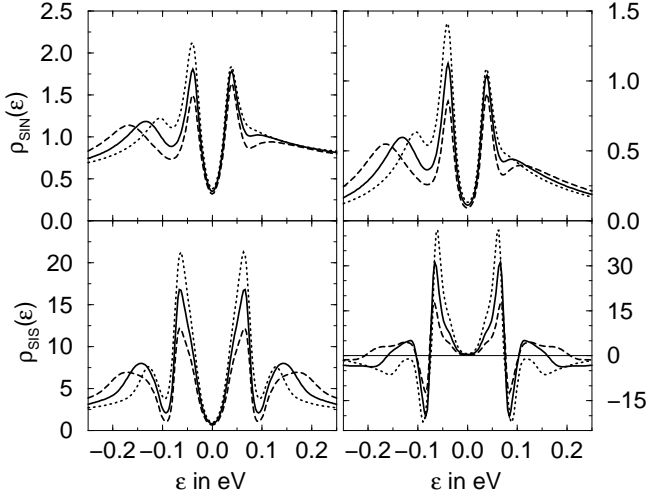}
\caption{ \label{fig4}
Differential tunneling conductance for SIN (top) and 
SIS (bottom) tunnel junctions for
$T=40$K.  Units are $eM_i^2$ for SIN and $2e^2T_i^2$ for SIS.
Left for isotropic tunneling ($i$=0),
right for coherent tunneling ($i$=1). 
Curves are for $g$=0.39 eV (dotted), 0.65 eV
(full line), and 0.90 eV (dashed).
}
\end{figure}
\noindent
In both cases, we 
observe a clear asymmetry with a dip-hump structure on
the negative bias side and a very weak feature on the positive side of the
spectrum, as in experiment \cite{Renner95}. 
The pronounced asymmetry is a result of the shallow band near the
$M$ point, $\xi_M \sim -\Omega $, which
enhances the coupling to the resonance mode for populated states.
Note that the hump position is strongly dependent on the coupling constant
in contrast to the position of the dip minimum.
The asymmetry in the peak height on either side of the
spectrum is sensitive to the coupling constant too.
In weak coupling the negative bias peak is higher due to the Van Hove
singularity at the $M$ point.
For stronger coupling the pronounced dip 
at negative bias reduces the height of the 
coherence peak on this side and shifts the hump to higher energies.
For $g$=$0.65$ eV (full lines in Fig.~4) the peaks at positive and negative
bias have roughly the same height, as in experiment \cite{Renner95}.

For an SIS junction, the single particle tunneling current is given in
terms of the spectral functions by
\begin{eqnarray}
I(V)=2e\sum_{\vec{k} \vec{p}} |T_{\vec{k}\vec{p}}|^2
\int_{-\infty}^\infty \frac{d\epsilon }{2\pi }
A(\vec{k},\epsilon ) A(\vec{p}, \epsilon+eV) \nonumber \\
\times \left\{ f(\epsilon ) - f(\epsilon + eV) \right\}
\end{eqnarray}
Again we show results for incoherent tunneling ($|T_{\vec{k}\vec{p}}|^2=T_0^2$)
and for coherent tunneling
with conserved parallel momentum,
$|T_{\vec{k}\vec{p}}|^2=\frac{1}{16}T_1^2 ( \cos k_x - \cos k_y )^4
\delta(\vec{k}_{||}-\vec{p}_{||})$.\cite{Chak93}
We show the SIS 
$dI/dV$ in the bottom panels of Fig.~4.
Our theoretical SIS curves for incoherent tunneling resemble very closely the
experimental results for BSCCO,\cite{Mandrus91}
unlike for coherent tunneling which exhibits
negative $dI/dV$ regions due to the strong anisotropy of $T_{\vec{k}\vec{p}}$.
Note that the dip-hump feature is strong on both sides
for an SIS junction in contrast to the SIN results.

In conclusion, we have shown that the momentum dispersion of the 
ARPES spectra, as detailed in recent experiments, can be
explained by a simple model which has as components A) a flat band
region near the chemical potential in the normal state dispersion near
the $(\pi,0)$ point of the zone; B) a nearly dispersionless
bosonic mode which is peaked in momentum near the $(\pi,\pi)$ point,
and which interacts with the fermionic degrees of freedom.
The theoretical tunneling spectra obtained with the same parameter set
are consistent with the experimental findings of an asymmetry of
the peak-dip-hump structure 
in SIN tunneling spectra.

The authors would like to thank A. Kaminski and J.C. Campuzano for
discussions concerning their photoemission data, and J. Zasadzinski 
concerning tunneling data.
This work was supported by the U.S. Dept. of Energy, Basic Energy Sciences,
under Contract No.~W-31-109-ENG-38.

\end{document}